\documentclass[letterpaper, 10 pt, conference]{ieeeconf}  

\IEEEoverridecommandlockouts                              
\usepackage{epsfig} 
\usepackage{amsmath,amssymb,amsfonts}
\usepackage{color}
\usepackage{booktabs,balance}
\usepackage{multirow}
\usepackage{mathtools}
\usepackage{pdfrender}

\usepackage{centernot}
\usepackage{color}
\usepackage[acronym]{glossaries}

\makeglossaries
\newacronym{LP}{LP}{Linear Programming}
\newacronym{MPC}{MPC}{Model Predictive Control}
\newacronym{QP}{QP}{Quadratic Programming}
\newacronym{LTI}{LTI}{Linear Time Invariant}
\newacronym{LQR}{LQR}{Linear Quadratic Regulator}
\newacronym{ARE}{ARE}{Algebraic Riccati Equation}
\newacronym{CLF}{CLF}{Control Lyapunov Function}
\newacronym{QCLF}{QCLF}{Quadratic Control Lyapunov Function}
\newacronym{PQCLF}{PQCLF}{Piecewise Quadratic Control Lyapunov Function}
\newacronym{CSCLF}{CSLF}{Constraints-Shaped Lyapunov Function}
\newacronym{MISO}{MISO}{Multiple Input - Single Output}
\newacronym{SISO}{SISO}{Single Input - Single Output}
\newacronym{MIMO}{MIMO}{Multiple Input - Multiple Output}
\newacronym{EAS}{EAS}{Euler Auxiliary System}
\newacronym{DA}{DA}{Domain of Attraction}
\newacronym{psd}{p.s.d.}{positive semi-definite}

\graphicspath{{Pictures/}}

\newcommand{\norm}[1]{\left\lVert#1\right\rVert}

\newcommand{\R}{\mathbb{R}}

\newcommand{\N}{\mathbb{N}}

\newcommand{\mc}[1]{\mathcal{#1}}

\newcommand{\continuanceref}{}

\newtheorem{theorem}{Theorem}

\newtheorem{definition}{Definition}
\newtheorem{proposition}{Proposition}

\newtheorem{example}{Example}
\newtheorem{continuancex}{Example \continuanceref}
\newtheorem{remark}{Remark}

\newtheorem{assumption}{Assumption}

\newenvironment{continuance}[1]
{\renewcommand\continuanceref{\ref{#1}}\continuancex[Cont'd]}
{\endcontinuancex}

\title{\LARGE \bf
On merging constraint and optimal control-Lyapunov functions
}

\author{Franco Blanchini \and Filippo Fabiani \and Sergio Grammatico
\thanks{F. Blanchini is with the Department of Mathematics and Informatics, University of Udine, Italy
	{\tt\small (franco.blanchini@uniud.it)}}%
\thanks{F. Fabiani is with the Department of Information Engineering, University of Pisa, Italy
	{\tt\small (filippo.fabiani@ing.unipi.it)}}%
\thanks{S. Grammatico is with the Delft Center for Systems and Control, TU Delft, The Netherlands
	{\tt\small (s.grammatico@tudelft.nl)}}%
}

\begin{document}

\maketitle
\thispagestyle{empty}
\pagestyle{empty}

\begin{abstract}
Merging two \glspl{CLF} means creating a single ``new-born'' \gls{CLF} by starting from two parents functions. Specifically, given a ``father'' function,  shaped by the state constraints, and a ``mother'' function, designed with some optimality criterion, the merging \gls{CLF} should be similar to the father close to the constraints and similar to the mother close to the origin. To successfully merge two CLFs, the control-sharing condition is crucial: the two functions must have a common control law that makes both 
Lyapunov derivatives simultaneously negative. Unfortunately, it is difficult to guarantee this property a-priori, i.e., while computing the two parents functions.
In this paper, we propose a technique to create a constraint-shaped ``father'' function that has the control-sharing property with the ``mother'' function.
To this end, we introduce a \emph{partial} control-sharing, namely, the control-sharing only in the regions where the constraints are active. 
We show that imposing partial control-sharing is a convex optimization problem. Finally, we show how to apply the partial control-sharing for merging constraint-shaped functions
and the Riccati-optimal functions, thus generating a \gls{CLF} with bounded complexity that solves the constrained linear-quadratic stabilization problem with local optimality.
\end{abstract}

\section{Introduction}
For solving constrained optimal-control problems, we need to face the following issue: in general, the cost-to-go function of the unconstrained problem is quite different from the one that shapes the constraints. An efficient solution can be achieved by combining the two functions via merging \cite{andrieu2010uniting,grammatico2014control}. 
Specifically, the merging function is a \gls{CLF} generated by two parent \glspl{CLF}, and represents an important trade-off \glspl{CLF} since, for instance, it may approximate the constraint-shaped function (father function) far from the origin, i.e, where the state constraints may be active, while being similar to the optimal one (mother function) close to the origin. However, there is a major issue in the merging procedure: although any pair of \glspl{CLF} can be successfully merged in dimension two \cite[Th. 1]{grammatico2014control}, this does not hold in higher dimensions. Remarkably, a crucial condition for merging two \glspl{CLF} is the control-sharing property, which is not necessarily satisfied in non-planar systems.

In this paper, we investigate a weaker property, hereby called \emph{partial} control-sharing, by considering a \gls{QCLF}, e.g.\ associated with the optimal \gls{LQR} for the unconstrained system, and a family of linear state constraints. We say that the quadratic function and the constraint functions have the partial control-sharing property if the \gls{QCLF} shares a control law with the constraint functions, provided that the latter are ``active''.

\subsection{Why merging?}
There are several approaches to deal with constrained optimal-control problems.
The most popular one is \gls{MPC} \cite{SznDam87,AllZhe00,GooSerDed06}, possibly in its explicit version \cite{BemMorDuaPis02}. While \gls{MPC} is powerful for discrete-time systems, it can become troublesome for continuous-time systems, as it requires fast sampling, hence long prediction horizons -- issues related to fast sampling can be partially accommodated via sub-optimal control approaches \cite{BlaMiaPel03}.

Perhaps the most popular approach is based on invariant sets and associated Lyapunov functions \cite{Ber72,GutHag85,HuLin01,BoyElgFerBal04,BlaMia15,HuTeeZac06}, where one faces the well-known trade-off between optimality and complexity by choosing among quadratic or non-quadratic functions (see \cite{HuLin01,BlaMia15,HuTeeZac06} for a more complete list of references). In this framework, constrained optimality can be tackled by means of gain-switching \cite{WreBel94}. Specifically,  an ``external guard" control is in charge to keep the state inside an invariant set (possibly the largest) compatible with the constraints. Next, this control is switched to the locally-optimal gain, as soon as the state reaches the largest constraint-compatible set \cite{GilTan91} of such a local regulator. The problem with this procedure is twofold: the high complexity of the representation of the sets involved and the discontinuity of the control law.

\subsection{Contribution}
In this paper, we aim at solving the constrained control problem with local optimality in continuous time. 
After formalizing the problem (\S \ref{sec:preliminaries}), the main contributions are:

\begin{itemize}
\item We provide necessary and sufficient conditions for the partial control-sharing property in the case of a \gls{QCLF}, $x^\top P x$, and a single linear constraint, $|f^\top  x| \leq 1$. We provide sufficient conditions
in the case of multiple constraints, $|f_i^\top  x| \leq 1$, for $i \in \{1, \ldots, s\}$. 

\item We verify the partial control-sharing in the region where $x^\top P x\leq \mu$ and $|f_i^\top  x| \leq 1$, for $i \in \{1, \ldots,s\}$, via convex programming (\S \ref{sec:partial_c-s}). By following a bisection procedure, one can find the largest $\mu$ for which the partial control-sharing property holds;

\item We derive the newborn \gls{CLF} by first smoothing the piecewise-quadratic function $\textrm{max}_i \{|f_i^\top  x|^2, x^\top P x\}$, and then by merging it with the optimal function, $x^\top P x$, with full control-sharing guarantee (\S \ref{sec:merging}). The resulting \gls{CLF} has a bounded complexity, being generated by the constraints and the optimal function.
\end{itemize}

\section{Problem formulation and preliminaries}\label{sec:preliminaries}
\subsubsection*{Notation} 
$\R$, $\R_{> 0}$ and $\R_{\geq 0}$ denote the set of real, positive real, non-negative real numbers, respectively. $\N$ denotes the set of natural numbers. 
For any positive (semi)definite function $V:\R^n \rightarrow \R_{\geq 0}$ and $\mu > 0$, the $\mu$-sublevel set is denoted by $\mc{L}_{(V/\mu)} \coloneqq \left\{x \in \R^{n} \mid V(x) \leq \mu\right\}$.

\subsection{An illustrative example}

\begin{figure}[!t]
	\centering
	\includegraphics[width=.8\columnwidth]{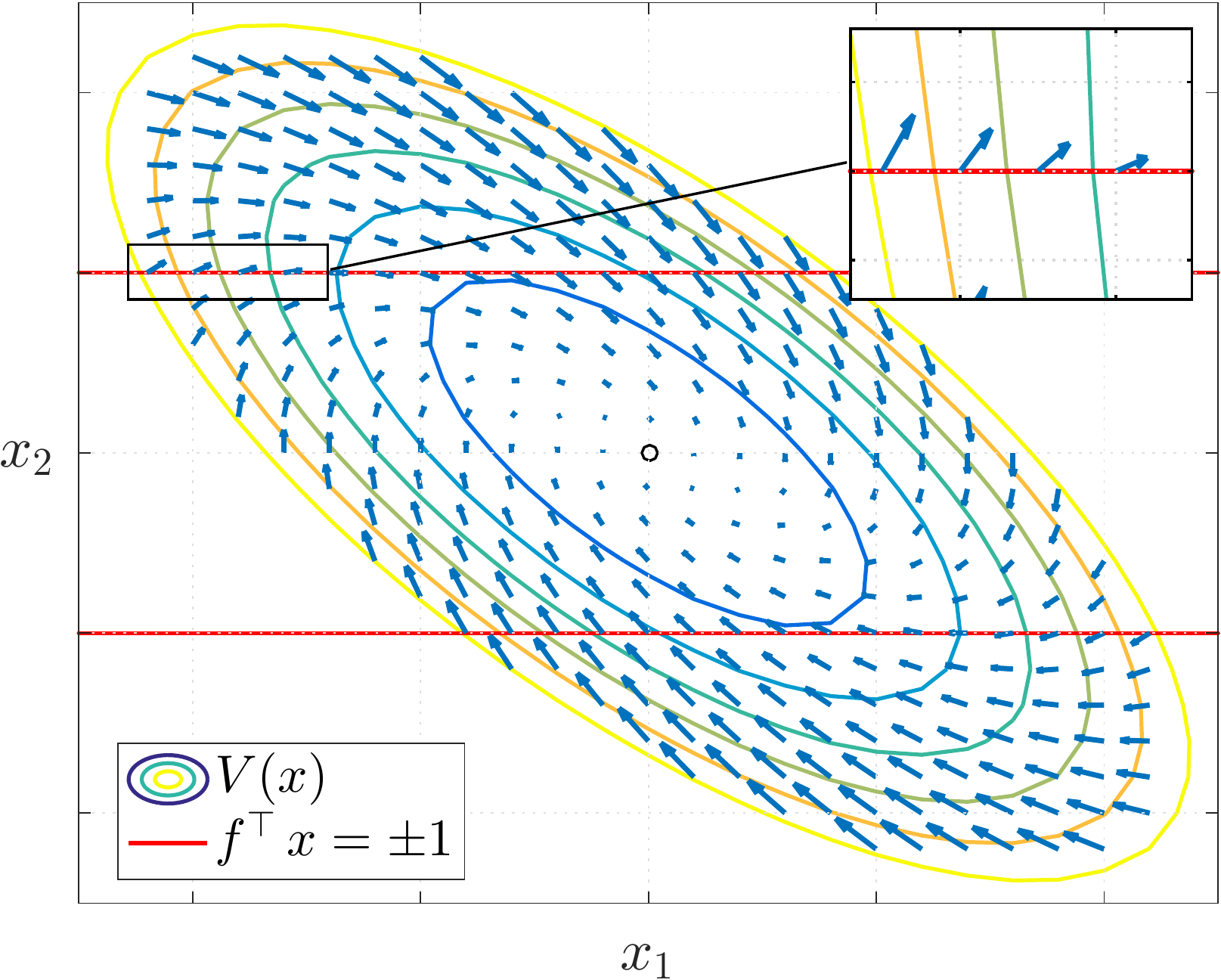}
	\caption{State behaviour of the pre-stabilized system in \eqref{eq:Ex1}. The blue arrows represent the derivative directions at every point inside the level curves of the associated \gls{QCLF} $V(x)$. The red lines denote the constraint on $x_2$.}
	\label{fig:DoubleInt}
\end{figure}

We start the paper with a simple, yet significant, example, to clarify the general problem addressed in the paper. 

\smallskip
\begin{example}\label{exa:ex1}Optimal constrained state feedback design.
\begin{equation}\label{eq:Ex1}
\begin{cases}
\dot{x} = \begin{bmatrix}
	0 & 1\\
	0 & 0
\end{bmatrix} x + \begin{bmatrix}
0\\
1
\end{bmatrix} u, \\ \\
y = \begin{bmatrix}
0 & 2
\end{bmatrix} x.
\end{cases}
\end{equation}

Let us consider the double integrator system in \eqref{eq:Ex1}, with performance output $y$, subject to linear constraint $\lvert y \rvert \leq 1$. The control input $u$ is preliminary chosen as an \gls{LQR} optimal feedback gain:
$u_0(x) = - R^{-1} B^\top P x = -(x_1 + \sqrt{2} \, x_2)$, where $P$ solves the classic \gls{ARE} with $Q = \left[\begin{smallmatrix}
1 & 0\\
0 & 0
\end{smallmatrix}\right]$ and $R = 1$.
We refer to this optimal control input as a pre-stabilizing compensator, which may fail when the constraint come into play. As shown in Fig.~\ref{fig:DoubleInt}, although the trajectories converge to the origin, there is a (symmetric) region close to the red boundaries where the optimal control drives the state outside the constraint. 
\hfill$\square$
\end{example}
\smallskip
In view of the previous example, throughout this paper we consider a generic \gls{LTI} system:
\begin{equation}\label{eq:OrigLTI}
\dot{x} = Ax + Bu,
\end{equation}
with state variable $x \in \R^{n}$, control input $u \in \R^m$, $A \in \R^{n \times n}$ and $B \in \R^{n \times m}$. As in Example~\ref{exa:ex1}, we suppose that the system in \eqref{eq:OrigLTI} is subject to linear constraints acting on the output variable. To tackle this problem, we also assume that the control $u$ may be chosen as the sum of two terms:
\begin{enumerate}
	\item a pre-stabilizing compensator $u_0(x) = -\hat{K} x$, $\hat{K} \in \R^{m \times n}$, that meets some optimality (local) conditions in absence of constraints;
	\item an additional control input $v = v(x) \in \R^m$, suitable to steer the system within the constraints.
\end{enumerate}
We aim at designing the additional control $v$ in order to ``enlarge'' the set of initial states that generates safe trajectories, while preserving  local optimality.

\subsection{Merging control Lyapunov functions: Background}

By referring to the linear system in \eqref{eq:OrigLTI}, in the following, we give some useful definitions.

\smallskip
\begin{definition}[Control Lyapunov Function]\label{def:CLF}
A positive definite, radially unbounded, smooth away from zero, function $V:\R^n \rightarrow \R_{\geq 0}$ is a control Lyapunov function (CLF) for \eqref{eq:OrigLTI} if there exists a locally bounded control law $u:\R^n \rightarrow \R^m$ such that, for all $x \in \R^n$, we have:
\begin{equation}\label{eq:CLF}
\nabla V(x) (Ax + Bu(x)) < 0.
\end{equation}
$V$ is a control Lyapunov function with domain $\mc{L}_{(V/\mu)}$, for $\mu > 0$, if \eqref{eq:CLF} holds for all $x \in \mc{L}_{(V/\mu)}$. 

Given some $\beta > 0$, the set $\mc{L}_{(V/\mu)}$ is $\beta$-contractive for \eqref{eq:OrigLTI} with control input $u(\cdot)$ if and only if:
\begin{equation*}
\nabla V(x) (Ax + Bu(x)) \leq - \beta V(x),
\end{equation*}
holds for all $x \in \mc{L}_{(V/\mu)}$.
\hfill$\square$
\end{definition}
\smallskip

\smallskip
\begin{definition}[Control-sharing property \textup{\cite[Def. 2]{grammatico2014control}}]
\label{def:C-S}
Two \glspl{CLF} $V_1$ and $V_2$ for \eqref{eq:OrigLTI} have the control-sharing property if there exists a locally bounded control law $u:\R^n \rightarrow \R^m$ such that, for all $x \in \R^n$, the following inequalities are simultaneously satisfied:
\begin{equation*}
\left\{ 
\begin{aligned}
	&\nabla V_1(x) (A x + B u(x)) < 0\\
	&\nabla V_2(x) (A x + B u(x)) < 0.
\end{aligned} 
\right.
\end{equation*}
\hfill$\square$
\end{definition}
\smallskip

\smallskip
\begin{definition}[Gradient-type merging \textup{\cite[Def. 3]{grammatico2014control}}]\label{def:M_CLF}
Let $V:\R^n \rightarrow \R_{\geq 0}$ be positive definite and smooth away from zero. $V$ is a gradient-type merging candidate if there exist two continuous functions $\gamma_1, \gamma_2 : \R^n \rightarrow \R_{\geq 0}$ such that $(\gamma_1(x), \gamma_2(x)) \neq (0, 0)$ and
\begin{equation*}
\nabla V(x) = \gamma_1(x) \nabla V_1(x) + \gamma_2(x) \nabla V_2(x).
\end{equation*}
$V$ is a gradient-type merging \gls{CLF} if it is also a \gls{CLF}.
\hfill$\square$
\end{definition}
\smallskip

In \cite{grammatico2014control}, a solution to the constrained control problem with local optimality is based on the following steps:
\begin{enumerate}
\item[S1)] \textit{Mother function}: Find the optimal \gls{QCLF}, $x^\top P x$, for the unconstrained system;
\item[S2)] \textit{Father function}: Find a constraint-shaped \gls{CLF}, e.g.\ by computing or approximating the largest controlled-invariant set;
\item[S3)] \textit{Merging}: Derive a \gls{CLF} that is similar to the father close to the constraints and to the mother near the origin.
\end{enumerate}

The third step is critical for two reasons. First, the possibility to merge two functions requires the control-sharing property \cite[Th. 2]{grammatico2014control}.
Unless we are dealing with a planar system, for which any two \glspl{CLF} share a control \cite[Th. 1]{grammatico2014control}, the control-sharing property may be not satisfied. 
Second, the high complexity of the maximal invariant set, i.e., the representation of the father function, might be inherited by the final merging function, which complicates the on-line computation of the control inputs. We face both problems by investigating a different condition, namely the partial control-sharing property.

\subsection{Problem formulation: Partial control-sharing}
We consider a region of bounded complexity of representation, which is shaped by the optimal and the constraint functions.
Then, let us consider the following assumption, which guarantees that the Riccati-optimal control, with infinite-horizon quadratic performance cost $J \coloneqq \int_0^\infty \|x\|^2_{Q} + \|u\|^2_{R} \, dt$, where $R \succ 0$ and $Q \succcurlyeq 0$, is stabilizing.

\smallskip
\begin{assumption}\label{ass:1}
The pair $(A, B)$ in \eqref{eq:OrigLTI} is controllable and the pair $(A, Q)$ is observable.
\hfill$\square$
\end{assumption}
\smallskip

We also assume that the state variable is subject to $s$ linear constraints, given by $|f_i^\top x| \leq 1$, for all $i \in \{1, \ldots, s\}$. By rearranging $f_i$ into the matrix $F \coloneqq [f_1, \ldots, f_s]^\top \in \R^{s \times n}$, we characterize the admissible state space as
\begin{equation*}
\mathcal{F}  \coloneqq \{x \in \R^n \mid \norm{F x}_{\infty} \leq 1\}.
\end{equation*}
For each constraint, we also introduce the functions $\psi_i:\R^n \rightarrow \R_{\geq 0}$, defined as $\psi_i(x) \coloneqq |f_i^\top x|^2$, so that $\mathcal{F}$ is characterized by the inequality
\begin{equation}\label{consfunct}
\Psi(x) \coloneqq \underset{i \in  \{1, \ldots, s\}}{\textrm{max}} \psi_i(x) \leq 1.
\end{equation}  

The optimal control gain matrix is $\hat{K} = R^{-1}B^\top P$ where $P \in \R^{n \times n}$ is the  
solution of the \gls{ARE}, $A^\top P + PA -PBR^{-1}B^\top P + Q =0$, and $V(x) = x^\top P x$ is the optimal unconstrained cost-to-go function (positive definite in view of Assumption~\ref{ass:1}).
Then, we shape the working region based on on $V$ and the constraints, i.e.,
\begin{equation*}
\mathcal{G}_\mu  \coloneqq \mathcal{F}  \cap \mc{L}_{(V/\mu)}.
\end{equation*}
The following definition limits the requirement of control-sharing
only when the boundaries are active.

\smallskip
\begin{definition}[$(\alpha, \beta)$-partial control-sharing property] 
Let $\alpha, \beta > 0$ be given.
The functions $V$ and $\Psi$ have the
$(\alpha, \beta)$-partial control-sharing property if there exists a locally-bounded control law $u: \R^{n} \rightarrow \R^m$ such that, for all $x \in \mathcal{G}_\mu$ and $i$ s.t $f_i^\top = \pm 1$, the following inequalities simultaneously hold:
\begin{equation}\label{partial}
\left\{ 
\begin{aligned}
&\nabla \psi_i(x) (A x + B u(x)) \leq -\alpha\\
&\nabla V(x) (A x + B u(x)) \leq -\beta \, V(x).\\
\end{aligned} 
\right.
\end{equation}\hfill$\square$
\end{definition}
\smallskip

\begin{remark}
We note that, if the partial control-sharing holds, then $\mathcal{G}_\mu$ is a control-invariant set. This type of regions has been considered as candidate control-invariant sets, see \cite{hu2010non,Ode02}. However, we ask something stronger than control invariance, which however only requires that $\dot{\psi}_i(x) < 0$ when the $i$-th constraint is active.
Thus, we require that, with the \emph{same} control input that keeps the state inside the set, we also have $\dot V(x) < 0$ on the boundary. In view of the final merging, this condition will ensure the full control-sharing property between the constraint-shaped function and the optimal one.
\hfill$\square$
\end{remark}

\section{Partial control-sharing conditions}\label{sec:partial_c-s}
\begin{figure}[!t]
	\centering
	\includegraphics[width=.9\columnwidth]{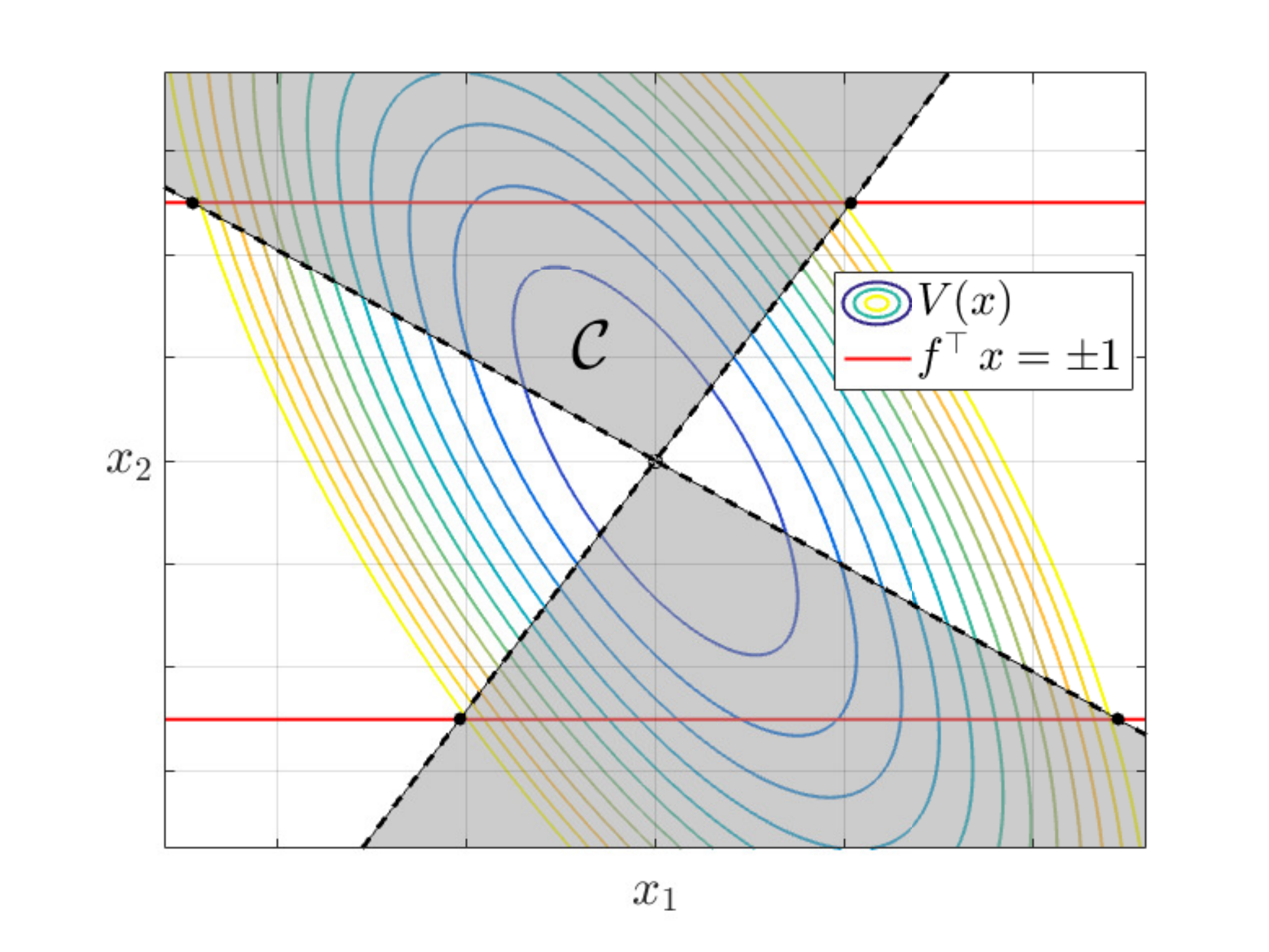}
	\caption{By referring to Example~\eqref{exa:ex1}, the shaded area represents the elliptical convex cone $\mathcal{C}$.}
	\label{fig:C_set}
\end{figure}

Without restrictions, we parametrize the control law as $u(x) = u_0(x) + v(x) = - \hat{K}x + v(x)$. Then, the system in \eqref{eq:OrigLTI} becomes:
\begin{equation}\label{eq:LTI}
\dot{x} = \hat{A} x + B v,
\end{equation}
with $\hat{A} \coloneqq (A - B \hat{K}) \in \R^{n \times n}$. We note that the optimal \gls{QCLF} $V(x)$ satisfies
\begin{equation}\label{eq:negder}
\dot{V}(x) = 2x^\top P \hat{A} x = -x^\top \hat Q x,
\end{equation}
with $\hat Q \coloneqq Q + PBR^{-1}B^\top P \succ 0$.

\subsection{MISO systems: Single state constraint}
First, we consider the case of a single constraint acting on the system in \eqref{eq:LTI}, i.e., $|f^\top x| \leq 1$. Then, let us define the following elliptical convex cone (an instance in Fig.~\ref{fig:C_set})
\begin{equation*}
\mc{C} \coloneqq \left\{ x/\lambda \in \R^n \mid x \in \mc{L}_{(V/\mu)} \cap \partial \mc{F} , \, \lambda > 0 \right\}.
\end{equation*}

Then, we have the following equivalence result.

\smallskip
\begin{theorem}\label{th:Th_MISO}
Let $V(x) = x^\top P x$ satisfy \eqref{eq:negder}, the function $\Psi(x) = \psi(x) = |f^\top x|^2$ be associated with the unique constraint, and let $\alpha, \beta, \mu > 0$ be given. 
The following statements are equivalent:
\begin{enumerate}
	\item[i)] $V$ and $\Psi$ have the $(\alpha,\beta)$-partial control-sharing property on $\mc{G}_\mu$;
	\item[ii)] $z^\top (\hat{Q} - \beta P) z - f^\top ( \hat{A} + \tfrac{\alpha}{2} I ) z \geq 0$ for all $z\in \mathcal{C}$, where $2 z^\top P B + f^\top B = 0$.
	\hfill$\square$
\end{enumerate}

\end{theorem}
\smallskip

\begin{proof}
\begin{figure}[!t]
\centering
\includegraphics[width=.7\columnwidth]{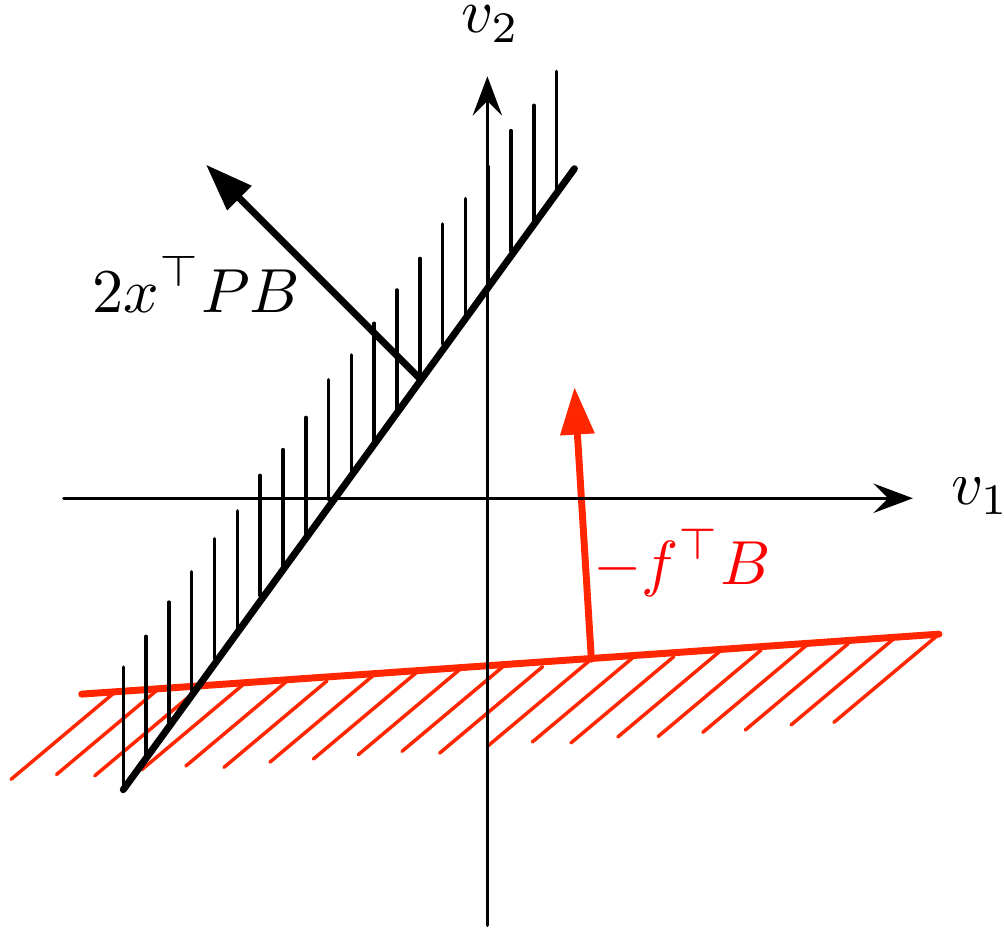}
\caption{Conditions in \eqref{eq:Hyperplanes1} in the case of two inputs ($m = 2$). The arrows represent the normal vectors to the hyperplanes.}
\label{fig:2DRegion}
\end{figure}
We consider the case $f^\top x =1$ only, as the proof for the symmetric one $f^\top x =-1$ is identical. Let $x \in \mc{G}_\mu$, and $\tilde{\alpha}=\alpha/2$. Then, the following conditions must hold:
\begin{equation*}\label{eq:Hyperplanes}
\left\{
\begin{aligned}
&f^\top (\hat{A}x+B v)  \leq -\tilde{\alpha} \\
&2 (x^\top P \hat{A} x + x^\top P B v) \leq - \beta x^\top P x, 
\end{aligned}
\right.
\end{equation*}
namely,
\begin{equation}\label{eq:Hyperplanes1}
\left\{
\begin{aligned}
& -f^\top B v \geq f^\top (\hat{A} + \tilde{\alpha} I) x\\
& 2 x^\top P B v \leq x^\top (\hat{Q} - \beta P) x.
\end{aligned} \right.
\end{equation}
These two inequalities are always satisfied if the vectors $-f^\top B$ and $2 x^\top P B$ are not aligned (see Fig.~\ref{fig:2DRegion}, that shows the situation with one constraint and $m = 2$). 
Hence, let us focus on the aligned case, i.e., when $2 x^\top P B + \lambda f^\top B = 0$ for some $\lambda > 0$. To guarantee the non-emptiness of the solution set in \eqref{eq:Hyperplanes1}, we must have that: 
\begin{equation*}
\begin{aligned}
\text{if} \quad & 2 x^\top P B + \lambda f^\top B = 0 \\
\text{then} \quad & x^\top (\hat{Q}-\beta P) x \geq \lambda f^\top (\hat{A} + \tilde{\alpha} I) x.
\end{aligned}
\end{equation*}
Thus, by dividing the first equality by $\lambda$ and both sides of the second inequality by $\lambda^2$, introducing the state transformation 
$z \coloneqq \left(\frac{x}{\lambda}\right) \in \mathcal{C}$, we obtain the desired condition.
\end{proof}
\smallskip

\begin{remark}
The tolerance $\beta>0$ can be small to make $(\hat{Q}-\beta P)$ positive definite\footnote{Precisely, $\beta$ must be smaller than the smallest eigenvalue of $\hat{Q}P^{-1}$.}. Thus, condition ii) in Theorem \ref{th:Th_MISO} can be checked via convex optimization by minimizing $z^\top (\hat{Q} - \beta P) z - f^\top ( \hat{A} + \tilde{\alpha}  I ) z$ on the convex domain $\mathcal{C}$ with linear constraint $2 z^\top P B + f^\top B = 0$.
\hfill$\square$
\end{remark}
\smallskip

For $\mu$ sufficiently small, we surely have feasibility. To enlarge the domain $\mc{G}_\mu$, we can progressively increase the parameter $\mu$ 
(i.e., consider larger level curves in $\mathcal{L}_{(V/\mu)}$) as long as  the condition  of the theorem is met, thus guaranteeing the existence of a common control law between $\Psi$ and $V$ with the largest $\mu$.

\begin{table}[!t]
	\caption{Optimal value of $z^\top (\hat{Q} - \beta P) z - f^\top ( \hat{A} + \tilde{\alpha}  I ) z$ for Example~\ref{exa:ex1}, with different parameter values.}
	\label{tab:ex1_val}
	\begin{center}
		\begin{tabular}{cccccccc}
			\toprule
			\multicolumn{1}{c}{$\alpha$} & \multicolumn{1}{c}{$\beta$}& \multicolumn{6}{c}{\centering $\mu$} \\
			 \cmidrule{3-8} & & \multicolumn{1}{c}{$0.5$} & \multicolumn{1}{c}{$0.75$} & \multicolumn{1}{c}{$1$} & \multicolumn{1}{c}{$1.5$} & \multicolumn{1}{c}{$3$} & \multicolumn{1}{c}{$10$} \\ \midrule
			\multirow{4}{*}{$0.001$} & $0.001$ & {43.57} & {11.22} & {6.43} & {3.62} & {1.78} & {0.71}\\
			\cmidrule{2-8} & $0.05$ & {41.96} & {10.76} & {6.13} & {3.42} & {1.64} & {0.61}\\
			\cmidrule{2-8} & $0.2$ & {37.15} & {9.36} & {5.24} & {2.83} & {1.24} & {0.33}\\
			\midrule
			\multirow{4}{*}{$0.1$} & $0.001$ & {43.17} & {11.04} & {6.30} & {3.54} & {1.73} & {0.69}\\
			\cmidrule{2-8} & $0.05$ & {41.56} & {10.58} & {6.01} & {3.34} & {1.59} & {0.59}\\
			\cmidrule{2-8} & $0.2$ & {36.75} & {9.18} & {5.12} & {2.75} & {1.19} & {0.30}\\
			\bottomrule
		\end{tabular}
	\end{center}
\end{table}

\begin{continuance}{exa:ex1}
	By applying the the conditions in \eqref{eq:Hyperplanes1} to $V(x)$ and $\psi(x) = x_2^2/4$ we obtain:
	\begin{equation*}
	\left\{
	\begin{aligned}
	&-2 v \geq -2 [x_1 + (\sqrt{2} - \tilde{\alpha}) x_2]\\
	&2(x_1  + \sqrt{2} x_2) v \leq (1-\sqrt{2})(x_1^2 + x_2^2) + (\sqrt{2} - 2\beta) x_1 x_2,
	\end{aligned}
	\right.
	\end{equation*}
	Thus, by introducing $\lambda > 0$ and following the same steps of the proof of Theorem~\ref{th:Th_MISO}, for $z \in \mathcal{C}$, if $z_1 + \sqrt{2} z_2 = - 1$, we must have
	\begin{equation*}
		(1 - \beta \sqrt{2}) (z_1^2 + z_2^2) + (\sqrt{2} - 2 \beta) z_1 z_2 \geq - 2[ z_1 + (\sqrt{2} - \tilde{\alpha}) z_2].
	\end{equation*}
	As summarized in Tab.~\ref{tab:ex1_val}, with small $\alpha$ and $\beta$, the latter condition is satisfied also for large values of $\mu$, guaranteeing the $(\alpha,\beta)$-partial sharing property between $\Psi$ and $V$ on $\mc{G}_\mu$. 
	
	\hfill$\square$
\end{continuance}

\subsection{MIMO systems: Multiple constraints}
Let us now consider the general case involving several state constraints. We must have that, whenever a set of constraints is active, i.e., $\psi_i(x) = 1$, the corresponding derivatives $\dot{\psi}_i$ and $\dot{V}$ shall be simultaneously negative by adopting the same control $v$.
Specifically, given any set of indices $K$, $H$ that denote active constraints, the $(\alpha,\beta)$-partial control-sharing property shall be ensured on each set:
\begin{multline*}
 \mc{A}_{K,H} \coloneqq  \left\{x \in \mc{L}_{(V/\mu)} \mid f_k^\top x = 1,\; f_h^\top x = -1, \right.\\
 \left. \textup{ for all } (k, h) \in K \times H \right\},
\end{multline*}
Let us restrict our investigation to the case in which all the constraints are equal to $1$; the other cases can be addressed by replacing $f$ by $-f$.
We call $\mc{A}$ the set of states where all $s$ constraints are active.

Before stating a sufficient condition for the partial control-sharing in MIMO systems, let us introduce the following set:
\begin{multline*}
\mc{V} \coloneqq \left\{ v \in \R^m \mid f_i^\top B v  \leq - f_i^\top (\hat{A} + \tfrac{\alpha}{2} I) x, \right. \\ 
\left. \textup{ for all } (x, i) \in  \mc{A} \times \{1, \ldots,s\} \right\}.
\end{multline*}

\smallskip
\begin{theorem}\label{th:Th_MIMO}
Under the same assumptions of Theorem~\ref{th:Th_MISO}, the functions $V$ and $\Psi$ have the $(\alpha , \beta )$-partial control-sharing property
if, for any set $\mc{A}_{K,H}$, it holds:
\begin{equation}\label{eq:Th_MIMO}
\underset{v \in \mc{V}}{\textrm{min}} \; \underset{x \in  \mc{A}_{K,H}}{\textrm{max}} \; x^\top (\beta P - \hat{Q}) x + 2 x^\top P B v \leq 0.
\end{equation}
\hfill$\square$
\end{theorem}

\smallskip

\begin{proof}
By construction, for any choice of active constraints in $K$ and $H$, when $x \in \mc{A}_{K,H}$, the conditions on the derivatives $\dot \psi_i(x) \leq -\alpha$ are satisfied for any $v \in \mc{V}$. Thus, the only concern refers to $V$. To ensure $\dot{V} < 0$, we must have $v \in \mc{V}$ such that
\begin{equation*}
	2 x^\top P \hat A x  + 2 x^\top P B v \leq -\beta x^\top P x,
\end{equation*}
which can be written as \eqref{eq:Th_MIMO}.
\end{proof}
\smallskip

\begin{figure}[!t]
\centering
\includegraphics[width=.7\columnwidth]{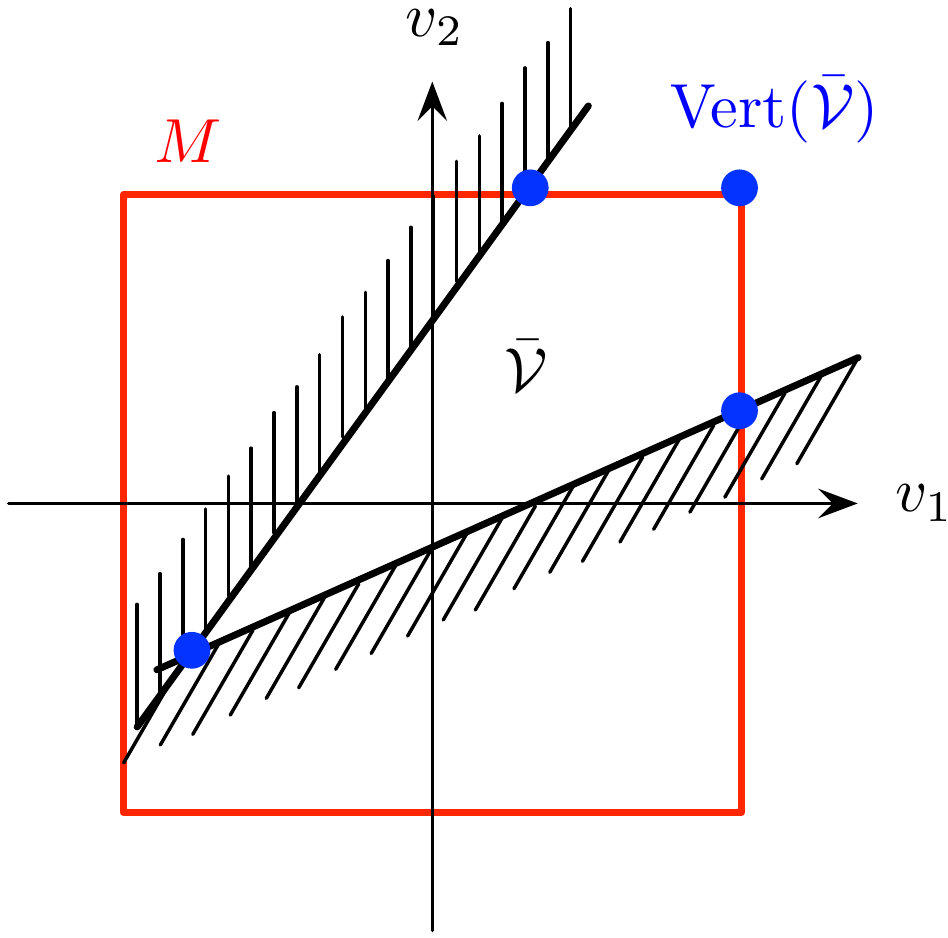}
\caption{Feasible set of the two dimensional \gls{LP} problem $\Phi$.}
\label{fig:VertReg}
\end{figure}

Here, $\beta$ shall be small enough to make $(\beta P - \hat{Q})$ negative definite.
For computational purposes, we may bound $v$ as $\| v \|_\infty \leq M$, with large $M$, and define the new set $\bar{\mc{V}}$ as
\begin{equation*}
\bar{\mc{V}} = \{v \in \R^m \mid \|v\|_\infty \leq M\} \cap	\mc{V}.
\end{equation*}

In that case, in view of \cite[Cor.~37.3.2]{rockafellar2009variational}, since $\bar{\mc{V}}$ and  $\mc{A}_{K,H}$ are two compact and convex sets and the function in \eqref{eq:Th_MIMO} is concave in $x$ and convex in $v$, we can exchange ``\textrm{min}'' and ``\textrm{max}''. Moreover, 
$$\Phi(x) \coloneqq \underset{v \in \bar{\mc{V}}}{\textrm{min}} \; x^\top (\beta P - \hat{Q}) x + 2 x^\top P B v$$
is an \gls{LP} problem on the compact set $\bar{\mc{V}}$. Then, if the feasible set is non-empty, an optimal solution does exist, and at least one these belongs to the set of vertices of the feasible region, namely $\textrm{Vert}(\bar{\mc{V}})$, as illustrated in Fig.~\ref{fig:VertReg}. Thus, we obtain that
\begin{equation*}\label{eq:Th_MIMO_minfirst}
\underset{x \in \mc{A}_{K,H}}{\textrm{max}}  \overbrace{\underset{v \in \textrm{Vert}(\bar{\mc{V}})}{\textrm{min}} \; x^\top (\beta P - \hat{Q}) x + 2 x^\top P B v}^{\eqqcolon \Phi(x)} \leq 0,
\end{equation*}
where $\Phi(x)$ is a concave function in $x$. As in the MISO case, the associated condition can be checked via convex optimization.
\hfill$\square$

\section{Application: smoothing and merging constraint and control-Lyapunov functions}\label{sec:merging}
In this section, we consider the problem of  shaping a \gls{CLF} 
starting from an optimal \gls{QCLF} and some constraint functions.
We first construct an intermediate function from $V$, suitably scaled by some $\mu$ that ensures partial control-sharing, and the constraint functions $\{ \psi_i \}_{i=1}^{s}$. Then, after a smoothing  procedure, we obtain a new \gls{CLF} that has the full control-sharing property with the optimal $V$.

\subsection{A smoothing method}
If there exists a control law such that $V$ and $\{\psi_i\}_{i = 1}^{s}$ simultaneously decrease along the solution to the system in \eqref{eq:LTI}, we can consider the following piecewise-quadratic candidate \gls{CLF}:
\begin{equation}\label{eq:PiecQuadLyapFun}
\hat{V}(x) \coloneqq \underset{i \in  \{1, \ldots, s\}}{\textrm{max}}\left\{V(x), \psi_i(x)\right\}.
\end{equation}

Since $\hat{V}$ is not a differentiable function, let us introduce the smoothed function, for some parameter $p \in \N$,
\begin{equation}\label{eq:SmoothedV}
V_p(x) \coloneqq \sqrt[p]{ \textstyle  V^p(x) +  \sum_{i = 1}^{s} \psi^{p}_i(x)}.
\end{equation}
 
In the following result, we show that for $p$ large enough, the function $V_p$ is a $\beta$-contractive \gls{CLF}.

\smallskip
\begin{proposition}\label{prop:betaEx}
Assume that $\hat{V}(x)$ is a $\beta$-contractive \gls{CLF} for \eqref{eq:LTI} with control law $v$. Then, there exists $\bar{p} \in \N$ and $\beta_p > 0$ such that, for all $p \geq \bar{p}$, $V^p_p(x)$ is a \gls{CLF} for \eqref{eq:LTI} with the same control law $v$.
\hfill$\square$
\end{proposition}
\smallskip

\begin{proof}
\begin{figure}[!t]
\centering
\includegraphics[width=0.8\columnwidth]{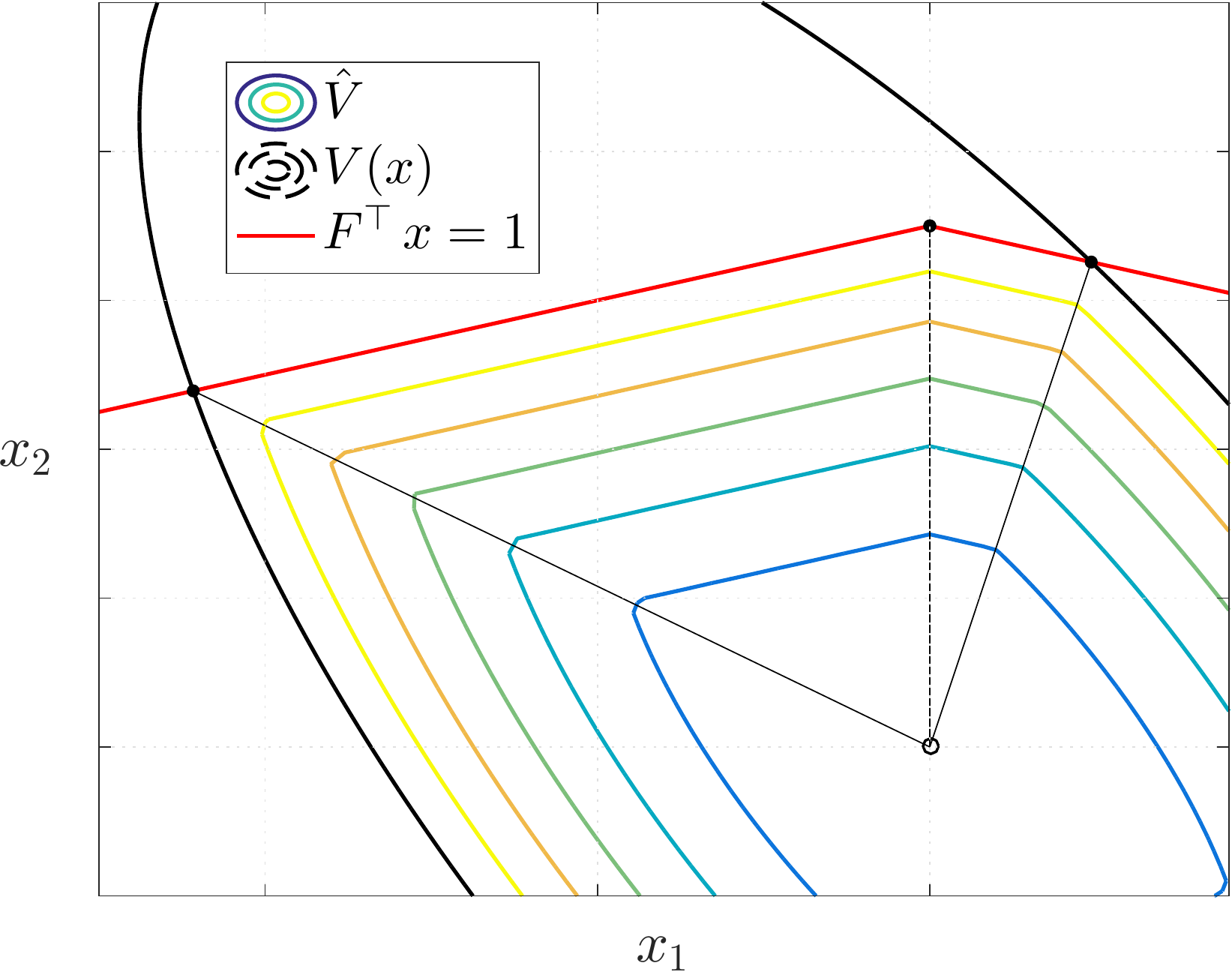}
\caption{Angular outline of $\hat{V}(x)$ (coloured level curves) inside the region bounded from the constraints (red lines) and $\partial\mathcal{L}_{V}$ (black dashed line). }
\label{fig:AngularHatV}
\end{figure}
Since $\hat{V}(x)$ is a piecewise quadratic candidate \gls{CLF}, there exists some $\hat{\beta} > 0$ such that $D^+ \hat{V}(x) \leq - \hat{\beta} \, \hat{V}(x)$, where $D^+$ denotes the upper-right Dini derivative. Then, let us define the Euler Auxiliary System (EAS) $x^+ \coloneqq x + \tau (\hat{A} x + B v)$, with $\tau > 0$ small enough. In view of \cite[Lemma 4.1]{blanchini1995nonquadratic}, there exists $\hat{\rho} \in \left[0, 1\right)$ such that, for the EAS, we have
$\hat{V}(x^+) \leq \hat{\rho} \, \hat{V}(x)$.
Without any restriction, the latter allows to consider an angular region that is bounded by the constraints and the \gls{QCLF} (the coloured level curves in Fig.~\ref{fig:AngularHatV}). 
Moreover, it follows from \cite[Th. 3.2]{blanchini1999new} that, for $V_p(x)$ in \eqref{eq:SmoothedV}, there exists some $\bar{p} \in \N$ and $\tilde{\rho} \in \left[0, 1\right)$ such that, for $p \geq \bar{p}$, $V_p(x^+) \leq \tilde{\rho} \, V_p(x)$.
Introducing two scale factors $\xi_i \in \left[0, 1\right)$, $i = 1, 2$, the idea is to enclose two level surfaces among the original bounded region $\mc{G}_{\mu}$ and the angular region previously introduced. As $p$ grows, such level curves approach the boundaries within which they are confined. Hence, the following chain of inequalities holds:
\begin{equation*}
\hat{V}(x) \leq V_p(x) \leq \frac{1}{\xi_1} V_p(x) \leq \frac{1}{\xi_2} \hat{V}(x),
\end{equation*}
which leads to 
$\partial\mathcal{L}_{\hat{V}} \supset \partial\mathcal{L}_{V_p} \supset \xi_1 \, \partial\mathcal{L}_{V_p} \supset \xi_2 \, \partial\mathcal{L}_{\hat{V}}$.
Then, the function $V_p(x)$ is $\rho_p$-contractive, with $\rho_p \coloneqq \hat{\rho}/\xi_2$, so $V_p(x^+) \leq \rho_p V_p(x)$. Directly from \cite[Lemma 4.2]{blanchini1999new}, with $v$, as $p \rightarrow \infty$, there exist a coefficient of contractivity $\beta_p \coloneqq (1 - \rho_p)/\tau$ such that $D^+ V_p(x) \leq - \beta_p V_p(x)$.
Consequently, since $V_p(x)$ is a positively homogeneous function, we have 
$D^+ V^p_p(x) \leq - \beta_p \, p \, V_p(x)$ as desired.
\end{proof}
\smallskip

\smallskip
\begin{proposition}\label{prop:sharing}
Let $V$ and $\Psi$ have the $(\alpha,\beta)$-partial control-sharing property. Then, for any $p \geq \bar{p}$, the functions  $V^p_p$ and $V$ have the full control-sharing property.
\hfill$\square$
\end{proposition}
\smallskip

\smallskip
\begin{proof}
By noticing that, if the optimal $V$ and the constraints have the $(\alpha,\beta)$-partial control-sharing property, the control law $v$ in Prop.~\ref{prop:betaEx} can be taken in such a way that $\dot V(x) \leq - \beta V(x)$, the proof directly follows from the results of the previous section.
\end{proof}

\subsection{A gradient-type merging: R-composition}
Once we have guaranteed the full control-sharing property between $V^p_p$ and $V$, we are in the position to achieve a successful merging. Next, we briefly recall the R-composition as a possible approach to merge two \glspl{CLF}, see \cite{balestrino2012new, balestrino2012multivariable,grammatico2013universal} for technical details. To obtain a merging function $V_{\wedge}$ that looks like $V$ close to the origin (locally optimal) and like the smoothed $V^p_p$ close to the constraints, the R-composition consists of the following steps:

\begin{itemize}

\item[R1)] Define $R_1, R_2: \R^n \rightarrow \R$, as $R_1(x) \coloneqq 1 - V_p^p(x)$ and $R_2(x) \coloneqq 1 - V(x)$;

\item[R2)] Fix $\phi > 0$, define the function $R_{\wedge} : \R^n \rightarrow \R$ (omitting the dependence on $\phi$) as
\begin{equation*}
R_{\wedge}(x) \!\coloneqq \!\rho(\phi) \!\left(\phi R_1(x) \!+\! R_2(x) \!-\! \sqrt{\phi^2 R_1^2(x) \!+\! R_2^2(x)} \right)
\end{equation*}
where $\rho(\phi) \coloneqq \left(\phi + 1 - \sqrt{\phi^2 + 1}\right)^{-1}$ is a normalization factor;

\item[R3)] Define the R-composition, $V_{\wedge} : \R^n \rightarrow \R_{\geq 0}$, as
\begin{equation*}
	V_{\wedge}(x) \coloneqq 1 - R_{\wedge}(x).
\end{equation*}

\end{itemize}

By computing the gradient $\nabla V_{\wedge}(x)$, it turns out from \cite[Prop.~5]{grammatico2014control} that $V_{\wedge}$ is a gradient-type merging candidate and can be used as a candidate \gls{CLF}.

\smallskip
\begin{table}[!t]
	\caption{Performance index $J$ for different values of $p$.}
	\label{tab:ex1_Jval}
	\begin{center}
		\begin{tabular}{ccccc}
			\toprule
			$p$ & $1$ & $2$ & $4$ & $30$\\ 
			\midrule
			$J$ & {82.95} & {24.95} & {23.13} & {27.17}\\
			\bottomrule
		\end{tabular}
	\end{center}
\end{table}
\begin{continuance}{exa:ex1}
\begin{figure}[!t]
	\centering
	\includegraphics[width=0.8\columnwidth]{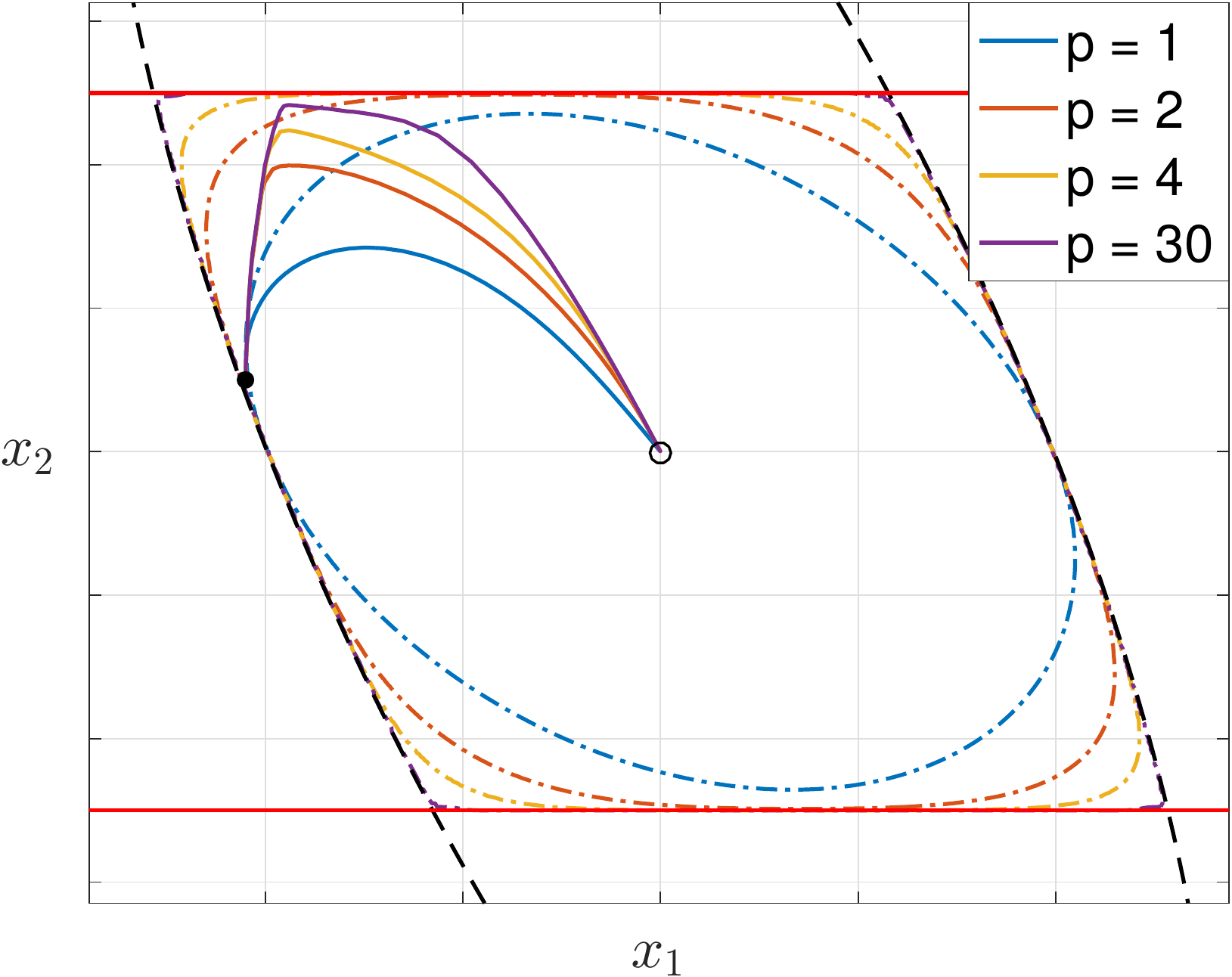}
	\caption{State behaviour (solid lines) of the system in \eqref{eq:Ex1} with gradient-type controller, for different values of $p$. The dashed-dotted lines corresponds to $V_{\wedge}(x) = \mu$, with $\mu = 1.4$, $\alpha = \beta = 0.1$, $\phi = 10$ and $x(0) = [-1.05, -0.1]^\top$.}
	\label{fig:comparison_p}
\end{figure}
Finally, we show an example of the correction made by gradient-based controller $v = - b^\top \nabla V_{\wedge}(x)$, with $V_{\wedge}$ obtained via the smoothing procedure and R-composition for different values of $p$. In Fig.~\ref{fig:comparison_p}, we shown controlled state trajectories, where the additional control input $v$ forces the state to remain inside the feasible region, $\mc{L}_{(V/\mu)}$, providing the values for the performance index $J$ in Tab.~\ref{tab:ex1_Jval}.
\hfill$\square$
\end{continuance}

\section{Conclusion and outlook}

Merging constraint functions and (locally) optimal control Lyapunov functions is key to design low-complexity (sub-) optimal control for constrained linear systems. 
Partial control-sharing is a promising approach for merging constraint and control-Lyapunov functions, under mild assumptions that can be checked via convex optimization.

Future research will investigate necessary and sufficient conditions for partial control-sharing in the presence of multiple state constraints. Control input constraints shall be considered as well. We shall also investigate sub-optimality bounds of certain merging procedures.

\balance
\bibliographystyle{IEEEtran}

\end{document}